
\input harvmac.tex
\hfuzz 15pt
\input amssym.def
\input amssym.tex
\input epsf\def\tfig#1{{
\xdef#1{Fig.\thinspace\the\figno}}Fig.\thinspace\the\figno
\global\advance\figno by1}


\input epsf

%



\def\a{\alpha}
\def\b{\beta}

\def\e{\epsilon}

\def\l{\lambda}

\def\s{\sigma}

\def\G{\Gamma}

\def\L{\lambda_{_{^L}} }



 %

\def\[{\left[}
\def\]{\right]}
\def\({\left(}
\def\){\right)}
\def\<{\left\langle\,}
\def\>{\,\right\rangle}


\def\inv{^{-1}}

 \def\frac#1#2{ {{\textstyle{#1\over#2}}}}
\def\inv{^{\raise.15ex\hbox{${\scriptscriptstyle -}$}\kern-.05em 1}}

 \def\IP{\relax{\rm I\kern-.18em P}}


%
\def\rb{ \noindent $\bullet$\ \ }

\def\zb{\beta}
\def\zs{\sigma}

\def\zG{\Gamma}
\def\la{\langle} \def\ra{\rangle}

\def\IZ{{ \Bbb Z} }

\def\dC{C\kern-6.5pt I}

\def\bsi{\bar{\zs}}

%


\def\LM{  \l_{_{M}} }

\def\tF{\tilde{F}}

\def\tL{\tilde \L}
\def\tM{\tilde \LM}

\chardef\tempcat=\the\catcode`\@ \catcode`\@=11
\def\cyracc{\def\u##1{\if \i##1\accent"24 i%
    \else \accent"24 ##1\fi }}
\newfam\cyrfam



\def\lmp#1#2#3{{Lett. Math. Phys.} {\bf #1} (#2) #3}

\def\hepth#1{{hep-th/}#1}

\def\encadremath#1{\vbox{\hrule\hbox{\vrule\kern8pt\vbox{\kern8pt
 \hbox{$\displaystyle #1$}\kern8pt}
 \kern8pt\vrule}\hrule}}

\def\hC{\hat{C}}


%
\lref\Lew{D.C. Lewellen, Nucl. Phys. {\bf B 372} (1992) 654.}
\lref\BPPZ{R. Behrend, P. Pearce, V.B. Petkova and J-B. Zuber,
Nucl. Phys. {\bf B 579} (2000) 707, hep-th/9908036.
}
 \lref\ValyaJB{V.B. Petkova and J-B. Zuber,
 Nucl. Phys. {\bf B 603} (2001) 449, hep-th/0101151.  }
\lref\AO{A. Ocneanu, Paths on Coxeter diagrams: From
Platonic solids and
singularities to minimal models and subfactors,
in {\sl Lectures on Operator Theory, }
 {Fields Institute, Waterloo, Ontario, April 26--30, 1995, }{(Notes taken by
S. Goto)},
{Fields Institute Monographies, AMS 1999,}{ Rajarama Bhat et al, eds.}}
\lref\BSz{ G. B\"ohm and K. Szlach\'anyi, \lmp{200}{1996}{437}, q-alg/9509008.}
\lref\FFFS{
G. Felder, J. Fr\"ohlich, J. Fuchs and  C. Schweigert, 
J. Geom. Phys. {\bf 34} (2000) 162, arXiv:hep-th/9909030; 
Compos. Math. {\bf 131} (2002) 189, hep-th/9912239.}
%
%
\lref\Run{I. Runkel,
Nucl. Phys. {\bf B 549} (1999) 563,  hep-th/9811178.}
 \lref\PTa{B.~Ponsot and  J.~Teschner,
Comm. Math. Phys. {\bf 224}, 3 (2001) 613,
math.QA/0007097.}
\lref\PT{B.~Ponsot and J.~Teschner,
  Nucl.~Phys.~{\bf B 622} (2002) 309, \hepth{0110244}.  }
 \lref\Pons{B.~Ponsot,  Recent progress in Liouville field theory, hep-th/0301193.}
\lref\Tes{J. Teschner, Remarks on Liouville theory with boundary,
hep-th/0009138.}
   \lref\FZZ{V.~Fateev, A.~B.~Zamolodchikov and A.~B.~Zamolodchikov,
Boundary Liouville field theory. I: Boundary state and boundary
two-point function,
hep-th/0001012.}
\lref\ZZPseudo{
A.~B. Zamolodchikov and A.~B. Zamolodchikov, ``Liouville field theory
on a
  pseudosphere,
  \hepth{0101152}. }
\lref\KPS{ I.~K.~Kostov, B.~Ponsot and D.~Serban,
{Nucl.\ Phys.}\ {\bf B 683}, 309
(2004), hep-th/0307189.  }
\lref\Alex{ S.~Y.~Alexandrov and  E.~Imeroni, Nucl.Phys. {\bf B 731} (2005) 242,
hep-th/0504199.}
\lref\BM{ A.~Basu and E.~J.~Martinec,
   Phys.\ Rev.  {\bf D} {\bf 72}, 106007 (2005),
\hepth{0509142}.}
\lref\KPa{ I. Kostov and V. Petkova,
Nucl. Phys.  {\bf B 770} [FS] (2007) 273,
 hep-th/0512346. }
 \lref\KPb{ I. Kostov and V. Petkova,
Nucl. Phys. {\bf B 769} [FS]  (2007)  175, hep-th/0609020. }
 \lref\BelZ{V. Pokrovsky, A. Belavin and Al.  Zamolodchikov, 
 Moduli integrals, ground ring and four-point function in minimal Liouville gravity, 
 in {\it Polyakov's string: Twenty five years after}, hep-th/0510214. }
 \lref\KostovGR{I.~K.~Kostov,
Nucl.\ Phys.\  {\bf B 689}, 3 (2004),
hep-th/0312301. }
\lref\Witten{ E.~Witten,
Nucl.\ Phys.\ {\bf B  373}, 187 (1992),
hep-th/9108004.}
 \lref\bershkut{
  M. Bershadsky and D. Kutasov,
Nucl. Phys. {\bf B 382} (1992) 213, \hepth{9204049}.  }
\lref\Hos{J.-E. Bourgine, K. Hosomichi, I. Kostov and  Y. Matsuo,  
Nucl.Phys. {\bf B 795} (2008) 243,  arXiv:0709.3912 [hep-th].}
\lref\FGP{P. Furlan, A.Ch. Ganchev and V.B. Petkova,
Int. J. Mod. Phys. {\bf A5} (1990) 2721.}

\lref\KR{A.N. Kirillov and N.Yu. Reshetikhin,
  Adv. Series in Math. Phys.  {\bf 7} (1989), 285.}
 \lref\SS{
 N. Seiberg and D. Shih, 
JHEP 0402 (2004) 021, hep-th/0312170.}
\lref\H{K. Hosomichi, Minimal open strings, arXiv:0804.4721 [hep-th].
}

\overfullrule=0pt
\Title{\vbox{\baselineskip12pt\hbox
{}\hbox{}}}
{\vbox{\centerline
 {Non-critical string  pentagon equations}
 \bigskip
 \centerline{ and their solutions}
 \vskip2pt
}}

  \centerline{P. Furlan$^{a,b}$, V.B. Petkova$^{c}$ and M. Stanishkov$^{c}$ }

 \vskip 1cm

 \centerline{ \vbox{\baselineskip12pt\hbox
{\it $^{a)}$Dipartimento  di Fisica Teorica
 dell'Universit\`{a} di Trieste, Italy,}
 }}
\centerline{ \vbox{\baselineskip12pt\hbox
{\it $^{ b)}$Istituto Nazionale di Fisica Nucleare (INFN),
Sezione di Trieste, Italy,}
}}
\vskip 5pt
\medskip
\bigskip
   \centerline{ \vbox{\baselineskip12pt\hbox
{\it  $^{c)}$Institute for Nuclear Research and Nuclear Energy (INRNE), }
}}
 \centerline{ \vbox{\baselineskip12pt\hbox
 {\it Bulgarian Academy of Sciences (BAS), Bulgaria}
 }}


\vskip 1.5cm

\baselineskip=11pt

\noindent
We derive pentagon type relations for the 3-point boundary tachyon correlation functions in the
non-critical open  string theory with generic $c_{\rm matter} <1$  and study
 their solutions in the case of FZZ branes. A new general formula for the Liouville 3-point factor  is derived.

\Date{}
\vfill
\eject

\baselineskip=14pt plus 1pt minus 1pt
\newsec{Introduction}

The associativity of the  operator product expansion (OPE) of the boundary fields  implies
an equation \Lew\  for the  boundary  3-point functions. It can be rewritten \BPPZ\  as a
pentagon type relation  for the boundary OPE coefficients,
similar to the pentagon relation for the fusing matrix, the quantum 6j-symbols.
The two equations   are 
identified in the rational case
\ValyaJB\
as part of the Big Pentagon relations
of a weak Hopf algebra \AO, \BSz, interpreted as the quantum symmetry of the given
BCFT.
The
 boundary field OPE coefficients play the role of the quantum 3j-symbols of this algebra.
In the  CFT described by diagonal modular invariants the two pentagon relations admit an identical form and thus the quantum 3j- and 6j symbols coincide up to a gauge  \BPPZ,  \FFFS, confirming an earlier result  in  \Run, where   the 3-point boundary functions were computed explicitly
in the $sl(2)$  case.
 The generalisation of the  quantum  6j symbols  to  the non-compact
 Liouville theory was found in \PTa, and the  boundary OPE coefficients   with boundaries of FZZ type \FZZ\  were described   in  \PT. The   gauge choice
 is  correlated  with the Lagrangean formulation of the theory, namely, the
 expression
 computed with the
 boundary $c>25$ Coulomb gas technique of \FZZ\ 
  is recovered
 as a residue from the integral formula in \PT.  The pentagon equations in this case were further discussed and used in \KPS.

 In this paper\foot{The paper is an extended version (section 3.2 is new) of a contribution to the proceedings of the 7-th International Conference
{\bf Lie Theory and Its Applications in Physics}, 18-24 June 2007, Varna,
ed. V.K. Dobrev et al,  (Heron Press, Sofia, 2008)  pp. 89-98.}
 we consider the non-critical string analog of the boundary pentagon relations and their solutions.  The theory combines  two Virasoro theories,  $c<1$ (matter) and $c>25$ (Liouville),
   so that the overall central charge is compensated by the central charge of a pair of free ghost fields.
 As in the bulk  \KPa,  \BelZ, the   emphasis is on the presence of non-trivial matter interaction
implemented conventionally by   the two $c<1$ screening charges.
  Our derivation  here exploits only  the factorisation of the 3-point tachyon boundary correlators
  into matter and Liouville factors.
It  yields the equations which can be obtained alternatively  in the ground ring approach \Witten,
\bershkut, \KostovGR, \BM, \KPa, 
 using the  coefficients in  the OPE of the ground ring generators and the tachyons.
The result
 is a generalisation of the trivial matter case considered  in \KPS,  in which the
  tachyon  correlators are  described by the correlators in the pure Liouville theory but with additional constraints on the set of representations arising from the mass-shell condition.

  The solution of the general equations is a product of the matter and Liouville 3-point boundary
  coefficients. We  consider
  the case when the matter fields are restricted by a  charge conservation condition with two types of screening charges, or/and correspond to degenerate $c<1$ Virasoro representations. In this case  
  the matter factor is given by the Coulomb gas expression and  in the non-rational case  can be recovered by analytic continuation of   the $c>25$ Liouville Coulomb gas expression.
 The Liouville factor of the tachyon 3-point correlator is given in principle by the integral  Ponsot-Teschner  (PT) formula.
We derive  instead a simpler compact  expression   using recursively the Liouville pentagon equations.   It is  valid for  charges corresponding to degenerate matter representations and is expressed in terms of  finite sum basic hypergeometric functions of type ${}_4\Phi_3$ with bounds  generically consistent  with the matter fusion rules. The formula   generalises a special (thermal) case  result of \Alex. Unlike the expression
found in \Alex\ our formula is
 explicitly
  invariant under cyclic permutations of the boundary fields.

  Furthermore  in Appendix B
we write down  the equations for the 3-point tachyon boundary correlators in the nonstandard
variant of the Liouville gravity  introduced in \KPa, \KPb. These equations  are  obtained  exploiting compositions of the general boundary ground ring OPE relations derived 
in \KPa.

 \newsec{Pentagon equations}

We shall keep only  the Liouville field labels for the tachyon boundary operator $T_{\b }^{(\e)}={}^{(\bsi_2, \s_2)}\, T_{(e,\b)}{}^{(\bsi_1, \s_1)}$  of chirality
$\e=\pm 1$,  $(e=\e\b-\e b^{\e}, \b)$,
while  the matter representation label $e$ and matter boundary  labels $\bsi_i$ will be   suppressed. The parameter
$b$ determining the central charges $c=13+6(b^2+1/b^2) >25$  and $c =13-6(b^2+1/b^2) <1$
of the two Virasoro theories  is  generically an arbitrary real number; most of the formulae below remain true for the rational (minimal matter) case. 
 It is convenient to use the  "leg factor"  normalisation
  \eqn\leg{
T_{\b }^{(\e)}(x)= 
 \Gamma(b^{\e}(Q-2\b))\ {\bf  c}(x) \,  e^{2i \e (\b-b^{\e})\chi(x)} \, e^{2\b\phi(x)}\,.
 }
The scaling dimensions
are
given respectively by
\eqn\dms{\eqalign{
&\triangle_L(\b)=\b(Q-\b)\,,  \ \  \ Q= 1/b+b\,, \cr
&\triangle_M(e)=e(e-e_0)
\,,  \ \ \  e_0 = 1/b-b\,, \cr
&\triangle_M(e)+\triangle_L(\e e+ b^{\e})=1=-\triangle_{\rm ghost\,{\bf c}}\,.
}}

\rb  The pentagon relations take simple recursive form when one of the
 operators corresponds to a  fundamental  degenerate Virasoro representation.
Starting with the
Liouville case, one has (see e.g., \KPS)
\eqna\pent
$$\eqalignno{
&C^L_{\zs_3\,, \zb_2 - t{b\over
2}}\left[\matrix{\zb_2&-{b\over 2}\cr
\zs_4& \s_2
} \right]\,
C^L_{\s_2 =\zs_3\pm {b\over 2}
\,,\zb_3}\left[\matrix{\zb_2-t{b\over 2}&\beta_1\cr
\zs_4&\sigma_1} \right] = \cr
& & \pent {} \cr
&F^L_{+t}\left[\matrix{\zb_2&-{b\over 2}\cr
\zb_3&\beta_1} \right]\,
C^L_{\zs_3 \pm {b\over 2}\, \beta_1-{b\over
2}}\left[\matrix{-{b\over 2}&\beta_1\cr
\zs_3&\sigma_1}  \right]\,
C^L_{\zs_3 \,,\zb_3}
\left[\matrix{\zb_2&\beta_1-{b\over 2}\cr
\zs_4&\sigma_1} \right]  +\cr
& {}\cr
&F^L_{-t}\left[\matrix{\zb_2&-{b\over 2}\cr
\zb_3&\beta_1} \right]\,
C^L_{\zs_3\pm {b\over 2}\, \beta_1+{b\over
2}}\left[\matrix{-{b\over 2}&\beta_1\cr
\zs_3&\sigma_1} \right]\,
C^L_{\zs_3 \,,\zb_3}\left[\matrix{\zb_2&\beta_1+{b\over 2}\cr
\zs_4&\sigma_1} \right]\,, \qquad t=\pm 1\,.
}$$
 The relation of the OPE coefficients to the  (cyclically symmetric) 3-point  correlators is
\eqn\refle{\eqalign{
C^L_{\zs_2 \,,Q-\zb_3}
\left[\matrix{\zb_2&\beta_1\cr
\zs_3&\sigma_1} \right] =\la {}^{\s_1}B_{\zb_3}{}^{\zs_3}\, B_{\beta_2}{}^{\zs_2}
B_{\beta_1}{}^{\zs_1}\ra &= {{}^L}C_{\b_3,\b_2,\b_1}^{\s_3,\s_2,\s_1}\cr
&= S(\s_1,\b_3, \s_3)\,
 {{}^L}C_{Q-\b_3,\b_2,\b_1}^{\s_3,\s_2,\s_1}\,
}}
where $ S(\s_1,\b_3, \s_3)$ is the reflection amplitude \FZZ.
The Coulomb gas constants computed for labels $\{\b_i \}$  restricted by the
  charge conservation condition  $\sum_i \b_i-Q=-mb -\frac{n}{b}$ (or
any reflection of this condition) are recovered as residues from the expression
in \PT.  

Similarly the matter pentagon equation  reads
\eqna\pentm
$$\eqalignno{
&C^M_{\bsi_3\,, e_2 + t'{b\over
2}}\left[\matrix{e_2&{b\over 2}\cr
\bsi_4& \bsi_2
} \right]\,
C^M_{\bsi_2 =\bsi_3\mp {b\over 2}
\,,e_3}\left[\matrix{e_2+t'{b\over 2}&e_1\cr
\bsi_4&\bsi_1} \right] = \cr
& & \pentm {} \cr
&F^M_{+,-t'}\left[\matrix{e_2&{b\over 2}\cr
e_3&e_1} \right]\,
C^M_{\bsi_3 \mp {b\over 2}\, e_1-{b\over
2}}\left[\matrix{{b\over 2}&e_1\cr
\bsi_3&\bsi_1}  \right]\,
C^M_{\bsi_3 \,,e_3}
\left[\matrix{e_2&e_1-{b\over 2}\cr
\bsi_4&\bsi_1} \right]  +\cr
& {}\cr
&F^M_{-,-t'}\left[\matrix{e_2&{b\over 2}\cr
e_3&e_1} \right]\,
C^M_{\bsi_3\mp {b\over 2}\, e_1+{b\over
2}}\left[\matrix{{b\over 2}&e_1\cr
\bsi_3&\bsi_1} \right]\,
C^M_{\bsi_3 \,,e_3}\left[\matrix{e_2&e_1+{b\over 2}\cr
\bsi_4&\bsi_1} \right]\,, \qquad t'=\pm 1\,,
}$$
and the matter  constants  will be normalised to be $1$ for $ e_1+e_2+(e_0-e_3)-e_0=0$.
\medskip

\rb The fusion matrix elements  and the boundary OPE constants in \pent{}, \pentm{}\
containing a fundamental Virasoro representation  are known constants, which are recalled in Appendix A.   The matter and Liouville fusion matrix elements are related by analytic continuation. E.g. for the  choice  of  the chiralities of the three fields as $(+,-,+)$
\eqn\mas{
\b_3=e_3+b\,,\  \b_2=-e_2+1/b\,, \ \b_1=e_1+b\,
}
one has
$F_{s,t}^M=F_{-s,t}^L\,, \ \
\tF_{s,t}^M = \tF_{-s,t}^L\,, $
which implies  the following identities 
\eqn\diffr{\eqalign{
F^L_{++}F^M_{--} - F^L_{+-} F^M_{-+}& =0=F^L_{-+}F^M_{+-} - F^L_{--} F^M_{++}\,,
\cr
- F^L_{+,+}\, F^M_{+,-} + F^L_{+,-}\, F^M_{+,+} & ={Q-2\b_1\over Q-2\b_2} =- F^L_{-,-}\, F^M_{-,+} + F^L_{-,+}\, F^M_{-,-} \,.
}}

Now we multiply the matter and Liouville  pentagon identities \pent{}\ and \pentm{}\ for the same
 fixed $t=t'$ - consistent with a tachyon of negative chirality  $(e_2+t \frac{b}{2}, \b_2-t \frac{b}{2} )$ in the l.h.s.
 On the other hand in the r.h.s. we get besides the two tachyon contributions also two mixed  terms,  inconsistent with the mass-shell condition. Due to the first of the identities in \diffr\
 these mixed terms are cancelled in the linear combination of the $t=+1$ and $t=-1$ product identities taken with relative minus sign. To compute this linear combination one has to take into account
 the second identity \diffr\ and one finally obtains  for the normalised  as in \leg\  tachyon OPE constants $\hat{C}$
\eqn\gro{\eqalign{
&\hC_{\s_2
 \,,\zb_3}\left[\matrix{\zb_2\!-\!{b\over 2}&\beta_1\cr
\zs_4&\sigma_1} \right] +\cr
&  \sqrt{\L\, \LM}\
c(\b_2) \
c^M_{(-\bar{\delta})}(\bsi_2\!=\!\bar{\s}_3\!-\!\bar{\delta} {b\over 2},e_2, \bsi_4) \ c^L_{(\delta)}(\sigma_2
=\zs_3\!+\!\delta {b\over 2},
\b_2, \sigma_4)\,  \hC_{\s_2
\,,\zb_3}\left[\matrix{\zb_2\!+\!{b\over 2}&\beta_1\cr
\zs_4&\sigma_1} \right] = \cr
&{} \cr
&-\sqrt{\LM}\,c^M_{(\bar{\delta})}(\bsi_3,e_1, \bsi_1) \, \hC_{\zs_3 \,,\zb_3}
\left[\matrix{\zb_2&\beta_1\!-\!{b\over 2}\cr
\zs_4&\sigma_1} \right]
-\sqrt{\L}\,c^L_{(-\delta)}(\sigma_3,\b_1, \sigma_1) \,  \hC_{\zs_3 \,,\zb_3}\left[\matrix{\zb_2&\beta_1\!+\!{b\over 2}\cr
\zs_4&\sigma_1} \right],
}}
where $\delta\,, \bar{\delta}=\pm 1\,, $
\eqn\cmcoef{\eqalign{
& c^L_{(\mp)}(\sigma_3, \b_1,\sigma_1)=
 {2\sin\pi b(\beta_1\!\mp (\sigma_1\!+\!\sigma_3\!-\!Q)\!-\!{b\over 2})\,
\sin\pi b(\beta_1\!\mp (\sigma_3\!-\!\sigma_1)\!-\!{b\over 2}) \over
\sin\pi b(Q-2\beta_1 )}\,, \cr
}}
\eqn\mcoef{\eqalign{
&c^M_{(\pm)}(\bsi_3, e_1, \bsi_1)
={2\sin\pi b(e_1\!\mp (\bsi_1\!+\!\bsi_3\!-\!e_0)\!+\!{b\over 2})\,
\sin\pi b(e_1\!\mp (\bsi_3\!-\!\bsi_1)\!+\!{b\over 2})\over
\sin\pi b(e_0-2e_1 )}\,.
}}
and
\eqn\fusec{
c(\b_2)=- {\sin \pi b(Q-2\b_2)\over  \sin \pi b(2\b_2)}\,.
}
The constants   $\L, \LM$ in \gro\
are the two bulk
coupling constants, following the notation in  \KPa.
Similarly one obtains the dual equation  with $\tL=\L^{1/b^2}\,, \tM=\LM^{-1/b^2}$
\eqn\grod{\eqalign{
&  - \sqrt{\tM}\,   \tilde{c}^M_{(\bar{\delta})}(\bsi_2,
e_2, \bsi_4) \,  \hC_{\s_2
\,,\zb_3}\left[\matrix{\zb_2\!-\!{1\over 2b}&\beta_1\cr
\zs_4&\sigma_1} \right] - \sqrt{\tL}\,\tilde{c}^L_{(\delta)}(\sigma_2, \b_2, \sigma_4)\,  \hC_{\s_2
\,,\zb_3}\left[\matrix{\zb_2\!+\!{1\over 2b}&\beta_1\cr
\zs_4&\sigma_1} \right]  \cr
&{}\cr
&=\hC_{\s_3 \,,\zb_3}\left[\matrix{\zb_2&\beta_1\!-\!{1\over 2b}\cr
\zs_4&\sigma_1} \right] +\cr
&   \sqrt{\tL\, \tM}\,
\tilde{c}(\b_1) \,
{\tilde c}^M_{(-\bar{\delta})}(\bsi_3\!=\!\bar{\s}_2\!-\!{\bar{\delta} \over 2b},e_1, \bsi_1) \, {\tilde c}^L_{(-\delta)}(\sigma_3\!=\!\zs_2\!-\!{\delta \over 2b},
\b_1, \sigma_1)\,  \hC_{\s_3
\,,\zb_3}\left[\matrix{\zb_2&\beta_1\!+\!{1\over 2b}\cr
\zs_4&\sigma_1} \right],
}}
replacing  the constants in  \cmcoef{}\  (and  \fusec{})
with their  duals, obtained by the change $b \to 1/b $
(for $\b_i $ - fixed), while the dual of the matter constant  \mcoef{}\ is obtained with $b\to -1/b$, so that
\eqn\dualc{\eqalign{
 \tilde{c}_{(\mp)}^M
(\bsi_3, e,\bsi_1)  
={2 \sin \pi{1\over b}( e\!-\!{1\over 2b}\mp(\bsi_3\!+\!\bsi_1\!-\!e_0))\,
 \sin \pi {1\over b}(e\!-\!{1\over 2b}\mp (\bsi_3\!-\!\bsi_1))\over
 \sin \pi {1\over b}(2e-e_0)}\,.
 }}

The two sets of equations \gro,\grod\  are precisely the  equations one obtains  starting from a 4-point function with a ground ring generator added
and then  inserting the  coefficients in  the  expansion  of the product of the ground ring generator
with the left or  right tachyons  (see formulae (A.36-A.38) of \KPa; the computation there completes
earlier  partial results  \refs{\bershkut,\KostovGR,\BM} for these OPE coefficients).

\subsec{\bf  Special case - trivial matter}

We choose as before the chiralities of type $(+-+)$. 
For  trivial matter, i.e., a charge conservation condition with no screening charges, 
\eqn\trivm{
e_0= e_1+(e_2+\frac{b}{2}) +(e_0-e_3)\equiv e_{12}^3 +e_0+ \frac{b}{2}\,  \ \Rightarrow \
\b_{23}^1 + \frac{b}{2}= Q
 }
the matter boundary 3-point functions are trivial and
the pure Liouville identity  \pent{}\   $(t=+1)$ (normalised with the leg factors) with  $\b_i$ restricted by \trivm\  simplifies
to
\eqn\purl{\eqalign{
&\hat{C}_{\zs_3\pm {b\over
2}\,,\zb_3}\left[\matrix{\zb_2-{b\over 2}&\beta_1\cr
\zs_4&\sigma_1} \right] = -\sqrt{\L}\,
 c_{(\mp)}^L(\sigma_3, \b_1,\sigma_1)\ \hat{C}_{\zs_3 \,,\zb_3}
\left[\matrix{\zb_2&\beta_1+{b\over 2}\cr
\zs_4&\sigma_1} \right]  \cr
&+F^L_{++}\ \Gamma(\frac{1}{b}(Q-2\b_2+b))\
 \Gamma(b(Q-2\b_1)) \, \Gamma(b(2\b_3-Q)\
 C^L_{\zs_3 \,,\zb_3}
\left[\matrix{\zb_2&\beta_1-{b\over 2}\cr
\zs_4&\sigma_1} \right]  \cr
&{}\cr
&=-\sqrt{\L}\,
 c_{(\mp)}^L(\sigma_3, \b_1,\sigma_1)\ \hat{C}^L_{\zs_3 \,,\zb_3}
\left[\matrix{\zb_2&\beta_1+{b\over 2}\cr
\zs_4&\sigma_1} \right] +  {2\pi G_2(\s_3,\b_2, \s_4)  \over  2\sin(\pi b(Q-2\b_1)) } \,.
}}
We have used that  for the values in \trivm\ $F^L_{++}$ has a zero $\sum_i \b_i^{'} -Q \rightarrow 0$, while the  Liouville reflected 3-point constant  has a singularity
with  residue $1/2\pi$.
Thus the  second term  in \purl\ reduces to the (leg-normalised)  reflection Liouville amplitude \FZZ,
 \eqn\Ltwop{\eqalign{
\Gamma(\frac{1}{b}(Q-2\b)) &\,\Gamma(b(Q-2\b))\,
S(\s_2,\b,\s_1)
={2 \pi\over Q-2\b}\,  G_2(\s_2,\b,\s_1)\,,
\cr
G_2(\s_2,\b,\s_1)&={\L^{{1\over 2b}(Q-2\b)}\,
S_b(2\b-Q) \over \prod_{s=\pm} \, S_b(\b +s(\s_2+\s_1-Q))\,
S_b(\b +s(\s_2-\s_1))}\,,\cr
&{}\cr
G_2(\s_2,\b,&\s_1)\, G_2(\s_2,Q-\b,\s_1)={S_b(2\b-Q) S_b(Q-2\b)}\,,
}}
where $S_b(x)\!=\!\G_b(x)/\G_b(Q-x)$ and   $\G_b(x)$ is   the double Gamma  function. The amplitude
 $G_2$ is the solution   of the equations
\eqn\twopoint{\eqalign{
-\sqrt{\L}\,  c_{(\mp)}^L(\sigma_3, \b_1,\sigma_1)\  G_2(\s_3,\b_1+ \frac{b}{2}, \s_1)&=G_2(\s_3\pm \frac{b}{2},\b_1, \s_1)\,,\cr
-\sqrt{\tL}\,  \tilde{c}_{(\mp)}^L(\sigma_3, \b_1,\sigma_1)\  G_2(\s_3,\b_1+ \frac{1}{2b}, \s_1)&=G_2(\s_3\pm \frac{1}{2b},\b_1, \s_1)\,.
}}
Since the general identity \gro\ should reduce for the values in \trivm\ to the simpler identity \purl,
this implies a restriction on the  unknown matter OPE coefficients  involved  in \gro.

The identity \purl\ acquires a more symmetric form when rewritten for the cyclically symmetric correlator
of type $(---)$ \foot{This  equation has been independently written down recently in \Hos.} obtained by two reflections \refle, now written for  the normalised correlators
 \eqn\reflec{\eqalign{
\hat{C}_{Q-\b_3,\b_2,\b_1}^{\s_4,\s_2,\s_1}&= {1\over b} {1\over  2 \sin(\pi b(2\b_3-Q))}\, G_2^{-1}(\s_1,\b_3, \s_4)\,\hat{C}_{\b_3,\b_2,\b_1}^{\s_4,\s_2,\s_1}  \cr
&
={1\over b^2}\, {\sin(\pi \frac{1}{b}(Q-2\b_1))\over \sin(\pi b(Q-2\b_3))}{G_2(\s_2,\b_1, \s_1)\over G_2(\s_1,\b_3, \s_4)}\,
\hat{C}_{\b_3,\b_2,Q-\b_1}^{\s_4,\s_2,\s_1} \,. 
}}

\noindent
 We give for comparison the equations analogous to \twopoint\   for the nontrivial  matter  2-point
 amplitude
 \eqn\twopointm{ \eqalign{
-\sqrt{\LM}\,  c_{(\mp)}^M(\bsi_3, e,\bsi_1)\  G_2^M(\bsi_3,e - \frac{b}{2}, \bsi_1) & =G_2^M(\bsi_3\pm \frac{b}{2},e, \bsi_1)\,,\cr
  - \sqrt{\tM}\,
 \, \tilde{c}^M_{(\mp)}(\bsi_3, e, \bsi_1) \, G_2^M(\bsi_3,e+\frac{1}{2b}, \bsi_1) & =  G_2^M(\bsi_3 \pm \frac{1}{2b},e, \bsi_1)\,.
}}

 \newsec{\bf  Solutions }

\subsec{\bf  The matter factor}

 We shall  start with the solution of the 2-point  equations \twopointm\ for
 the matter degenerate values $2e =mb-n/b=: 2e_{m,n}$,
 where $m,n$ are nonnegative integers, $m,n\in \IZ_{\ge 0}$.
The solution is  expressed  conveniently as
\eqn\soltwoma{\eqalign{
&G_2^M(\bsi_2, e, \bsi_1)= {(-1)^{(m+1)(n+1)}\ \LM^{2e-e_0\over 2b}\over S_b((m+2)b)\, S_b(\frac{n+2}{b})}\ {G_M(\bsi_1,e\!-\!b\!-\!(m\!+\!1)b,\bsi_2)\over G_M(\bsi_1,e\!-\!b\!+\!\frac{n+1}{b},\bsi_2) }\cr
&= \LM^{2e-e_0\over 2b} \L^{-(Q+m b+{n}/{b}\over 2b)} G_2(\bsi_2+b, -e-\frac{n}{b},\bsi_1+b)  {S_b(2Q+m b+\frac{n}{b}) S_b(\frac{1}{b})\over S_b^2(\frac{n+2}{b})}
}}
where
\eqn\mG{
G_M(\bsi_3, e_2,\bsi_2):= S_b(-e_2+\bsi_2+\bsi_3)\,  S_b(e_0\!-e_2+\bsi_3-\bsi_2)\,.
}
The representation of  \soltwoma\  in terms of $S_b(x)$  is
not unique, but the expression is finite for
the concrete  values $e=e_{m,n}$ and reduces to a finite product of
$\sin$'s. The equations \twopointm\  allow to extend the  formula  \soltwoma\
to $m=n=-1$
and furthermore to the degenerate values $2e =e_0 - (m+1)b +\frac{n+1}{b}$ with
$m,n\in \IZ_{\ge 0}$. \foot{Note that unlike the Liouville case the analytic continuation of the  two thermal cases $n=0$ or $m=0$  to  generic  values of $e$ leads to  different  results, effectively inverse to each other.}

\medskip

\rb 
The solution of the pair of  equations \gro,  \grod\  is given by a factorised expression combining the known Liouville expression \PT\
and a solution of the matter boundary pentagon equation.
The solution
of the matter boundary pentagon equation is a
 generalisation
to  generic $b^2$ of  the solution in the rational
$c<1$ case,  where the fusing matrix is given  \FGP\  by a product  of two basic ${}_4\Phi_3$ hypergeometric
functions known to  represent \KR\  the quantum
6j symbols. The change of gauge affects only the prefactor.
The non-rational generalisation  is possible   either if  the representations  are chosen to correspond to degenerate $c<1$ Virasoro representations, or, if
a charge conservation condition with integer numbers of matter screening charges is imposed: to both we refer as "Coulomb gas" cases.  The solutions in these cases are alternatively  reproduced starting from  the general formula of Ponsot and Teschner \PT. Thus to obtain the matter constant for
\eqn\coulg{
e_{123}-e_0\equiv e_1+e_2+e_3-e_0=mb-n/b\,, \ \ m,n\in \IZ_{\ge 0}
}
we start from the
Liouville Coulomb gas expression for $\a_{123}-Q=-mb-n/b$ derived  as a residue of the formula in \PT.  We
rewrite this particular solution of the Liouville pentagon equation \pent{}\  in terms of finite products of Gamma and $\sin$ functions and  then continue analytically
the result by  replacing $b^2\!\to\!-\!b^2\,,$ and  $\a_i b\!\to e_i b$. The final result is a solution of the matter
pentagon equation \pentm{}\ and can be again expressed in compact form  in terms of 
the ratios of double Gamma functions $\G_b(x)$ using the notation \mG,
\eqn\matcob{\eqalign{
&C^M_{\bsi_2 \,,e_0-e_3}\left[\matrix{e_2&e_1\cr
\bsi_3&\bsi_1} \right]= {{}^M}C_{e_3,e_2,e_1}^{\bsi_3,\bsi_2,\bsi_1}=
(-1)^{m+n  } \, \LM^{e_{123}-e_0\over 2b}\, \Pi_M(e_3,e_2,e_1)
\, \times \cr
& 
 { (-1)^{mn}S_b({b}\!+\!2e_1\!-\!{m}{b})\over S_b({b}\!+\!2e_1\!+\!\frac{n}{b})}  
\sum_{k=0}^m\sum_{p=0}^n
{G_M(\bsi_3, e_2\!-\!b\!-\!kb,\bsi_2)\, G_M(\bsi_3, e_0-e_3\!-\!b\!-\!\frac{n\!-\!p}{b} ,\bsi_1)  \over G_M(\bsi_3, e_0-e_3\!-b\!+\!(m\!-\!k)b,\bsi_1) G_M(\bsi_3, e_2\!-\!b\!+\!\frac{p}{b},\bsi_2)}
\,\times \cr
 &
 {S_b({b}\!+\!2e_3\!-\!(m\!-\!k)b)\, S_b({b}\!+\!2e_2\!-\!k b)\, S_b( \frac{1}{b}\!-\!2e_2\!-\!\frac{p}{b})\,
 S_b(\frac{1}{b}\!-\!2e_3\!\!-\!\frac{n\!-\!p}{b}) \over
 S_b((k+1)b) \,S_b((m\!-\!k\!+\!1)b)\,S_b(\frac{p\!+\!1}{b})\,S_b(\frac{n\!-\!p\!+\!1}{b}) }\,,
}}
where
\eqn\prefM{\eqalign{
 \Pi_M(e_3,e_2,e_1)&= 
{b^{Q(e_{123}-e_0)}\G_b(b) S_b(\frac{n\!+\!1}{b}) \over \G_b(\frac{1}{b} + e_0-e_{123}) }
\prod_i {\G_b(\frac{1}{b}-2e_i) S_b({b}\!+\!2e_i\!+\!\frac{n}{b})\over  \G_b({b} + 2e_i+ e_0-e_{123})}
=\cr
&{b^{Q(e_{123}-e_0)}\G_b(\frac{1}{b}) S_b(m\!+\!1)b) \over \G_b({b} - e_0+e_{123}) }
\prod_i {\G_b({b}+2e_i) S_b(\frac{1}{b}\!-\!2e_i\!+{m}{b})\over  \G_b(\frac{1}{b} - 2e_i- e_0+e_{123})}
\,.
}}
This formula is derived  for generic values of  $\{e_i\}$, subject of the constraint \coulg,
but  it reproduces as well the constants with degenerate values of $e_i$.

\subsec{\bf  The Liouville 3-point factor}

The matter charge conservation condition \coulg\ rewrites   as a relation for the Liouville labels, e.g., with the choice of chiralities ${(+-+)}$ one has
\eqn\coull{\b_{13}^2\equiv \b_1+\b_3-\b_2=(m +1)b -\frac{n}{b}\,.
}
 In  addition
 we choose also degenerate values for all matter labels, or equivalently, in terms of the Liouville
 labels $\b_i$ of the three fields in the correlator we  take 
 \eqn\dmrep{
 \b_i= b +m_i b  -\frac{n_i}{b}\,, \  2m_i, 2n_i\in \IZ_{\ge 0} \,, \ \ m,n\in \IZ_{\ge 0}\,.
 }
We further impose  the  (matter) fusion rule restriction that all $m_{ij}^k\,, n_{ij}^k\, , i\ne j\ne k\ne i$ are non-negative integers,  so that $\sum_{i=1}^32m_i=0$ mod $2$. Other possible choices 
correspond to Liouville reflections $Q-\b_i$ of some of the labels in \dmrep\ and the corresponding 3-point correlator is obtained with the help of the reflection relation \reflec.

For such values  of $\{\beta_i\}$ 
 the
 integral Ponsot - Teschner  formula for the Liouville 3-point boundary constant simplifies. Taking into account two infinite series of poles it rewrites
 as a sum of two terms, 
each  expressed in terms of a product of basic ${}_4\Phi_3$ hypergeometric functions, one given by a finite (of range $n$ -  as in \coull), the other - by an infinite, sum. A resummation of the infinite sums
was performed in \Alex\  in the particular  
case  $m_i=0\,, i=1,2,3$ of \dmrep.\foot{The formal resummation in \Alex, which  we believe is correct  only when applied to the sum of the two terms, amounts to a relation  for  ${}_3\Phi_2$ q- Saalschutz type functions.}

We shall follow here a different route to obtain a general simple formula without exploiting the integral
PT representation. 
Namely we shall use recursively the Liouville equations \pent{}\  
 starting
from the  simplest correlator with three identical fields $\b_i=b$
\eqn\spefa{\eqalign{
&\hat{C}_{b,b,b}^{\s_3,\s_2,\s_1} =  {2\pi  \, \sqrt{\L}^{-1}  \over \, g_-(\s_1,b/2,\s_2) }\ (G_2(\s_3,b,\s_1) - G_2(\s_3,b,\s_2) )\cr
& = {2\pi \, {\L}^{Q-3b \over 2b}\over S_b(\frac{2}{b}) }
{(\tilde{c}_1 (c_2-c_3)+\tilde{c}_2 (c_3-c_1)+\tilde{c}_3 (c_1-c_2))\over (c_2-c_1)(c_1-c_3)(c_3-c_2)}
}}
where
\eqn\dfcg{\eqalign{
&c_i=2\cos\pi b (b-2\s_i)\,,  \  \ \tilde{c}_i= 2 \cos \pi \frac{1}{b}(\frac{1}{b}- 2\s_i)\,,\cr
&g_{-}(\s_1,\b,\s_2)=2\cos \pi b (2\b-2\s_1) - 2\cos \pi b( b-2\s_2)\,
}}
and  recall that $\L^{1/2} \cos \pi b(Q-2\s)$
and $\tL^{1/2} \cos \pi  (Q-2\s)/b$ are   the boundary cosmological constant and its dual. 

The cyclic symmetry of the correlator is explicit in the second line of \spefa. This correlator,
originally  proposed in the microscopic approach of \KPS,   and  then  reproduced   in \Alex,
is itself  obtained directly from the (properly regularised) Liouville pentagon equation \pent{}\  for $t=1, \b_2=b=Q-\b_3 , \b_1=b/2$,   in a way similar to the derivation of the special case equation \purl.  
 For this choice of the parameters the l.h.s. and the first term in the r.h.s. of \pent{}\ are  identified with 
reflected trivial Coulomb gas correlators, so that they are represented by  2-point amplitudes - the ones appearing in \spefa. 

Let us first consider the 
"thermal" case with  
 all $n_i=0$ in \dmrep. The Liouville correlator in \spefa\
(normalised with the leg factors \leg) coincides with the tachyon correlator itself since it  corresponds to a  trivial matter condition with $m\!=\!0\!=\!n$ in \coulg, \coull.  Applying first  
trivial matter 
equations of the type in  \purl\  we  get the most general  correlator with $m_{13}^2=0$. Then   using the general equation \pent{}\ (for shifts of the pair $(\b_3,\b_2)$), we obtain, 
   denoting  $m=m_{13}^2\,, s= m_{12}^3$, 
\eqn\nderaa{\eqalign{
&
{}^LC_{\b_3,\b_2, \b_1}^{\s_3,\s_2 ,\s_1} =
\L^{-1/2}\, \Pi_L(\b_3,\b_2,\b_1)\,S_b(\frac{2}{b})\,
 \times \cr
 &{}\cr
& \Bigl(
 { G_2( \s_2\!+\!b ,\b_1,\s_1)\, \over g_-(\s_2,\b_1\!-\!\frac{b}{2},\s_1)}
 { S_b((s\! +\!1)b)\over S_b((s\!+\!m\!+\!1)b) }
 \sum_{p=0}^{s}  {S_b((p\!+\!m)b\!+\!Q)\over S_b(pb\!+\!Q)}
{G_2( \s_2\!+\!p\frac{b}{2},\b_2\!-\!p\frac{b}{2},\s_3) \over
G_2(\s_2\!+\!b\!+\!(p\!+\!m)\frac{b}{2},\b_1\!-\!(p\!+\!m)\frac{b}{2},\s_1)} \cr
&{}\cr
&+ {G_2( \s_1\!-\!b,\b_1,\s_2)\, \over 
g_-(\s_1,Q-\b_1\!+\!\frac{b}{2},\s_2)}
{S_b((m\!+\!1) b)\over S_b((s\!+\!m\!+\!1)b)}
 \sum_{r=0}^m {S_b((r\!+\!s)b\!+\!Q)\ \ \ \ G_2(\s_1\!-\!r\frac{b}{2} ,\b_3\!-\!r\frac{b}{2},\s_3)\over 
 S_b(rb\!+\!Q)\, G_2(\s_1\!-\!b\!-\!(r\!+\!s)\frac{b}{2},\b_1\!-\!(r\!+\!s)\frac{b}{2},\s_2)}\Bigr)
}}
where  
\eqn\prefL{
\Pi_L(\b_3,\b_2,\b_1)={b^{e_0(Q-\b_{123})}
\Gamma_b(2Q-\b_{123}) \Gamma_b(Q-\b_{23}^1) \Gamma_b(Q-\b_{13}^2) \Gamma_b(Q-\b_{12}^3)\over S_b(\frac{1}{b})S_b(\frac{2}{b})\, \G_b(Q) \G_b(Q-2\b_1) \G_b(Q-2\b_2)  \G_b(Q-2\b_3)}\,.
}
In   the overall prefactor in  the  product of the Liouville and matter  correlators  combining
 \prefM, \prefL\ and  the leg factor normalisation \leg\   the $\G_b$ functions are fully compensated, e.g. with the choice of the chiralities $(+,-,+)$ one has
 \eqn\prefful{\eqalign{
&\G(b(Q-2\b_3))\G(\frac{1}{b}(Q-2\b_2))\G(b(Q-2\b_1))\,  \Pi_M(\b_3\!-\!b,-\b_2\!+\!\frac{1}{b},\b_1\!-\!b)\Pi_L(\b_3,\b_2,\b_1) \cr
&
={2\pi \, S_b(2\b_1-b)\, S_b(2\b_2-\frac{1}{b})\, 
S_b(2\b_3 -b )  \over  S_b(2\b_1-b-m b)\, S_b(2\b_2-\frac{1}{b}-\frac{n}{b})\, S_b(2\b_3 -b -mb)
} {S_b(\frac{n\!+\!1}{b})\over S_b(\frac{1}{b})S_b(\frac{2}{b})}\,.
}}
In  Appendix C we give a few  explicit  examples demonstrating the two formulae  \matcob,   \nderaa. We shall rewrite now
\nderaa\  in a form which reveals its  symmetry under cyclic permutations. Let us first  introduce some general notation 
\eqn\ginstc{\eqalign{
& G^{(-)}(\s_2,\b,\s_1):  = S_b(-\b+\s_2+\s_1)S_b(Q-\b+\s_2-\s_1)  
= (G^{(+)}(\s_2,Q-\b,\s_1))^{-1}\,,\cr
&{G^{(\pm)}(\s_2,\b-\frac{b}{2},\s_1) \over
G^{(\pm)}(\s_2,\b +\frac{b}{2},\s_1)} =g_{\pm}(\s_2,\b,\s_1) = 2 \sin \pi b(Q-2\b)\ c_{\pm}^L(\s_2,\b,\s_1)
\,.
}}
For a non-negative  integer $k$ and an integer $n$ of parity $p(n)$
denote
\eqn\ants{
B(\s_2,\s_1)^{(k; p(n))}:=
{G^{(-)}(\s_2, -\frac{k b}{2} -\frac{n}{2b},\s_1)\over G^{(-)}(\s_2, b+\frac{kb}{2}-\frac{n}{2b},\s_1)}=
(-1)^{(k+1)(n+1)}B(\s_1,\s_2)^{(k; p(n))}
}
which is  expressed as a $k+1$ order polynomial in $\{c_i\}$ using that for $k\ne 0$
$$g_-(\s_2, \frac{b}{2} - k \frac{b}{2}+\frac{n}{2b}, \s_1) \  g_-(\s_2, \frac{b}{2} +k \frac{b}{2}+\frac{n}{2b}, \s_1)= c_1^2 +c_2^2 - c_1\,  c_2
(-1)^n 2\cos \pi k b^2  - (2 \sin \pi  k b^2)^2\,
$$
while $B(\s_2,\s_1)^{(0;p(n))}=(-1)^n c_2-c_3$.
Similarly one defines the dual $\tilde{B}(\s_2,\s_3)^{(n;p(k))}$ so that the reflection amplitude is expressed 
as the ratio
\eqn\antss{
{\L^{2 \b_2-Q\over 2b} G_2(\s_2,\b_2\!=\!b\!+\!m_2b\!-\!n_2/b, \s_3)\over  S_b(2\b_2-Q)}= 
{G^{(-)}(\s_2,\b_2,\s_3)\over   G^{(-)}(\s_2,Q-\b_2,\s_3)}=
{\tilde{B}(\s_2,\s_3)^{(2n_2;p(2m_2))}\over 
B(\s_2,\s_3)^{(2m_2; p(2n_2))}}\,.
}
Finally we introduce 
 \eqn\csympol{\eqalign{
&P_2\equiv P^{\s_3,\s_2,\s_1}_{\b_3,\b_2,\b_1}:=\cr
& (-1)^{m_{12}^3+2m_2}\,  \L^{-\frac{m_{12}^3}{2}}\, {S_b((2m_1+1)b) S_b((2m_2+1)b)\over S_b(b)}
  \sum_{p=0}^{m_{12}^3}  {  S_b((m_{12}^3+1)b)\over S_b((p+1)b)\, S_b((m_{12}^3+1-p)b)}\, \times
  \cr
 &\qquad \ \qquad {G_2( \s_2\!+\!p\frac{b}{2},\b_2\!-\!p\frac{b}{2},\s_3)\over G_2( \s_2,\b_2,\s_3)}{G_2( \s_2\!-\!(m_{12}^3\!-\!p) \frac{b}{2}, \b_1\!-\!(m_{12}^3\!-\!p) \frac{b}{2},\s_1)  \over 
G_2( \s_2, \b_1,\s_1)} \cr
& ={(-1)^{2m_2} S_b((m_{12}^3+1)b)\over S_b(b)}
\sum_{p=0}^{m_{12}^3}  {S_b(2m_1+(p-m_{12}^3+1)b) \,S_b(2m_2-(p-1)b)\over S_b((p+1)b \, S_b((m_{12}^3 +1-p)b)}\, 
 \times \cr 
&\qquad \ {G^{(-)}( \s_2,\b_2 -pb,\s_3) \over  G^{(-)}( \s_2,\b_2,\s_3) }
{ G^{(+)}( \s_2,\b_1-(m_{12}^3-p)b,\s_1)\over   G^{(+)}( \s_2 ,\b_1,\s_1)} \cr
}}
and similarly $P_1$ and $P_3$,  which can be obtained  from \csympol\ by cyclic permutations.
The finite sum \csympol\ is   proportional to a truncated ${}_4\Phi_3$ type function. It  can be  expanded as a polynomial in the variables $\{c_i\}$.

With this notation  
 \nderaa\  
is cast in a    form  generalising the second line  in \spefa,
\eqna\csymf
$$\eqalignno{
&
{}^LC_{\b_3,\b_2, \b_1}^{\s_3,\s_2 ,\s_1} 
=-{ \L^{Q-\b_{123}\over 2b}\, \Pi_L(\b_3,\b_2,\b_1) \over B(\s_1,\s_2)^{(2m_1;0)}\, B(\s_2,\s_3)^{(2m_2;0)}\, B(\s_3,\s_1)^{(2m_3;0)}}\, F_{\b_3,\b_2, \b_1}^{\s_3,\s_2 ,\s_1}\,,\cr
&{}\cr
&F_{\b_3,\b_2, \b_1}^{\s_3,\s_2 ,\s_1}=
(-1)^{2m_1} ((\!-1)^{2m_2}\tilde{c}_2\!-\! \tilde{c}_3)B(\s_3,\s_1)^{(2m_3;0)}P^{\s_3,\s_2,\s_1}_{\b_3,\b_2,\b_1}\cr 
&\qquad \qquad \ \ \qquad \qquad - (\!-1)^{2m_2} ((-1)^{2m_3}\tilde{c}_3\!-\!\tilde{c}_1)B(\s_2,\s_3)^{(2m_2;0)}P^{\s_2,\s_1,\s_3}_{\b_2,\b_1,\b_3}  & \csymf {} 
\cr
& {}   \cr 
&= 
\tilde{c}_1 B(\s_3,\s_2)^{(2m_2;0)}P^{\s_2,\s_1,\s_3}_{\b_2,\b_1,\b_3}+\tilde{c}_2  B(\s_1,\s_3)^{(2m_3;0)}P^{\s_3,\s_2,\s_1}_{\b_3,\b_2,\b_1}+
\tilde{c}_3 B(\s_2,\s_1)^{(2m_1;0)}P^{\s_1,\s_3,\s_2}_{\b_1,\b_3,\b_2}\,.
}$$
 In the
second equality of 
\csymf{}\
we have exploited the  relation
\eqn\csymrel{\eqalign{
 B(\s_3,\s_1)^{(2m_3;0)}P^{\s_3,\s_2,\s_1}_{\b_3,\b_2,\b_1}
 + {\rm cyclic \ permutations}   =  0
 }}
which  is proved using the alternative recursive derivations of \nderaa, i.e., the cyclic symmetry, now  explicit in \csymf.

  The composition of the reflection  of all three fields 
  with the reflection amplitude as in \refle\  and the duality transformation $b\to 1/b\,$ (changing notation
  $ m_i \to n_i$)  gives the correlator in the  other thermal 
  case when all $m_i=0$ in \dmrep. In that case  the product of $B^{(0;p(2n_i))}$  replaces the denominator 
  in  \csymf{}\ and the formula confirms the structure suggested in the microscopic approach of \KPS. The dual polynomial $\tilde{P}^{\s_3,\s_2,\s_1}_{\b_3,\b_2,\b_1}  $
 is  
  defined by  changing in \csympol\   $\b_i\to Q-\b_i\,, b \to 1/b \,, m_i  \to   n_i$. 
 With the help of some identities for the basic hypergeometric functions one reproduces  the  formula
found in \Alex\ for the  case $\{m_i=0\,, $  $n_i$ -  integers$\}$ \  by  exploiting in a formal way
the PT formula. The expression in \Alex\ is not explicitly symmetric under cyclic permutations, rather this symmetry is checked to hold on examples.
\medskip

\rb To obtain the Liouville correlator defined for the general values \dmrep\  
one can either use the dual pentagon equations, or, one  can start from  the  
correlator with all $m_i=0$.
In  one of the steps the special
case equation \purl\ has to be  extended so that
 the second term in the r.h.s. is given by $G_2$ times a non-trivial Coulomb gas Liouville correlator. The final result is an expression generalising  the first  line in
\csymf{},   
 \eqn\nderf{\eqalign{
&{}^LC_{\b_3,\b_2, \b_1}^{\s_3,\s_2 ,\s_1} =-{\L^{Q-\b_{123}\over 2b}\, \Pi'_L(\b_3,\b_2,\b_1) 
\over B(\s_1,\s_2)^{(2m_1;p(n_1))} B(\s_2,\s_3)^{(2m_2;p(n_2))}B(\s_3,\s_1)^{(2m_3;p(n_3))}}\, 
\times  \cr
&{}\cr
 & (-1)^{2m_2 2n_1}\Bigl((-1)^{2m_1+2n_2}
\tilde{B}(\s_2,\s_3)^{(2n_2; p(2m_2))} \tilde{P}^{\s_2,\s_1,\s_3}_{\b_2,\b_1,\b_3}\
 B(\s_3,\s_1)^{(2m_3; p(2n_3))} P^{\s_3,\s_2,\s_1}_{\b_3,\b_2,\b_1}\cr
&-
(-1)^{2m_2+2n_1}\tilde{B}(\s_3,\s_1)^{(2n_3; p(2m_3))}\tilde{P}^{\s_3,\s_2,\s_1}_{\b_3,\b_2,\b_1} 
\  
B(\s_2,\s_3)^{(2m_2;p(2n_2))} P^{\s_2,\s_1,\s_3}_{\b_2,\b_1,\b_3} 
 \Bigr)\,,\cr
 }}
  \eqn\prefl{
 \Pi'_L(\b_3,\b_2,\b_1) ={(-1)^{ m_{123}\, n_{123}\!+\!m_{123}\!+\!n_{123}}\, \Pi_L(\b_3,\b_2,\b_1) \ 
 S_b^3(\frac{1}{b})S_b(\frac{2}{b}-b)  
 \over
S_b(\frac{n_{12}^3+1}{b})S_b(\frac{n_{23}^1+1}{b})S_b(\frac{n_{13}^2+1}{b})S_b(\frac{n_{123}+2}{b}-b)}\,.
}
Here, say,  the polynomial $P_2$ is given by the first  formula \csympol,  where now all $\b_i$ are given by
\dmrep,  with only the sign in front of \csympol\  modified to  $(-1)^{m_{12}^3 (1+2n_3) +2m_3 2n_3+2m_2}=(-1)^{m_{123}2n_3 +2m_1}$. Let us also write down the expression for  one of the dual polynomials 
\eqn\dpol{\eqalign{
&  \tilde{P}_1\equiv  \tilde{P}_{\b_2, \b_1,\b_3}^{\s_2 ,\s_1,\s_3}= 
{(-1)^{n_{123}2m_2\!+\!2n_3}S_b(\frac{2n_1+1}{b}) S_b(\frac{2n_3+1}{b})
\over S_b(\frac{1}{b})}
 \sum_{u=0}^{n_{13}^2}   {S_b(\frac{n_{13}^2+1}{b})\over   S_b(\frac{1+u}{b})S_b(\frac{n_{13}^2+1-u}{b})}\times\cr
& 
{G_2(\s_1+\frac{u}{2b},Q-\b_1-\frac{u}{2b}, \s_2)\over G_2(\s_1,Q-\b_1, \s_2)}
 {G_2(\s_1-\frac{n_{13}^2-u}{2b},Q-\b_3-\frac{n_{13}^2-u}{2b},\s_3)\over G_2(\s_1,Q-\b_3,\s_3)}\,.
  }}
Formula \nderf\ gives the  general expression for the Liouville factor in  the tachyon 3-point boundary correlator with degenerate $c<1$ representations.  The cyclic symmetry of the full correlator
is ensured by construction and is equivalent to a relation generalising \csymrel, 
\eqn\gsymrel{\eqalign{
(-1)^{2n_2(2m_2+1)} B(\s_3,\s_1)^{(2m_3;p(2n_3))}\, P_2
 + {\rm cyclic \ permutations}   =  0\,
 }}
 and its dual with the dual polynomials and $m_i \leftrightarrow n_i$. 
 In particular when all $m_i=0$ the    dual relation  reproduces  the cyclic identity satisfied by
 the  first order dual polynomials $\tilde{B}(\s_2,\s_3)^{(0; p(2m_2))}=(-1)^{2m_2} \tilde{c}_2-\tilde{c}_3$, etc.,  which appear in the numerator in \csymf{}.
 The composition of duality transformation $b\to 1/b, m_i \leftrightarrow n_i$
 with reflection of all three fields keeps  \nderf{}\ invariant.
\medskip

\rb We conclude with some remarks.

The above solutions of the Liouville and matter equations  defined for generic values of the
parameters apply  in particular to the rational (minimal gravity) theory in which case there may appear further truncations of the sums.

The (thermal) matter 3-point function \matcob\  is 
given by the same basic hypergeometric function as one  of the polynomials $P_i$ in the  numerator of the Liouville factor with proper identification of the parameters 
\eqn\matli{\eqalign{
&
 {}^MC_{e_3,e_2,e_1}^{\bsi_3,\bsi_2,\bsi_1} 
 \sim  P^{\frac{1}{b}-\bsi_1,\frac{1}{b}-\bsi_3,\frac{1}{b}-\bsi_2}_{\frac{1}{b}-e_1,\, e_3+b,\, e_2+b} =P_3\,.
 }}
Similarly  for $ e_1=m_1b\,, e_3=m_3b\,, e_2=e_0-m_2b$ \matcob\ is identified with the polynomial $P_1$  in \csymf, etc.  
 Analogous  to \matli\ formulae hold for the 
 case $\b_i=b-n_i/b$,   relating \matcob\  to one of the  dual polynomials  with $\s_i=\bsi_i+b$.

The factorised matter - Liouville correlator
contains "too many" boundaries - their cardinality should be  the same as that of the
set of tachyons.  Examples of a  linearly independent set of boundaries is provided by the  "trivial matter boundaries", i.e.,  one $\s_i$ is set to zero,  while the intermediate two  are fixed by the fusion rules: the matter factor  is reduced to a  correlator of  chiral vertex operators.  On the level of 1-point functions or boundary states  one can represent 
 the states with general degenerate matter boundaries $\bsi_i$ as linear combinations of 
FZZ states with shifted boundary parameters $\s_i$ \SS.  It remains to 
look for some  lifting of 
this fusion type relation to the boundary correlators, see the recent work \H\ for a step in this direction.

 Another   choice,  the consistency of which deserves to be investigated,  
 are the "tachyonic boundaries"  when
the pairs $(\bsi_i,\s_i)$ themselves satisfy the  mass-shell condition required by BRST invariance.  Such correlators  could be rather  interpreted as  the
"string q-6j symbols", i.e., the OPE coefficients of the string CVO. They satisfy the pair of equations \gro,\grod, with 
 correlated signs $\delta,\bar{\delta}$,  preserving the chosen 
(chirality) type of the relation.  For such a "tachyonic" choice the  matter boundary parameters are to be extended to the complex values $2\bsi_i   = e_0 \pm iP$ corresponding to the 
FZZ branes $2\s_i =Q - i P$. 

\bigskip
\noindent
{\bf Acknowledgements}

\noindent
We thank Ivan Kostov for useful discussions.
P.F. acknowledges the support of the Italian Ministry of University and Research (MIUR).
V.B.P. acknowledges the hospitality of Service de Physique
Th\'eorique, CEA-Saclay, France,  ICTP and INFN, Italy.
This research is supported in part by the European Community
through   MCRTN ForcesUniverse, contract MRTN-CT-2004-005104 and by the
French-Bulgarian project RILA, contract  3/8-2006.

\appendix{A}{\bf  Data on fundamental OPE coefficients}

 The fusing matrix elements  and the boundary OPE constants in \pent{}, \pentm{}\
containing a fundamental $c>25$ or $c<1$  Virasoro representation   are known Coulomb gas constants,
e.g.,
\eqn\fga{\eqalign{
&F_{s,t}^L=
F^L_{\beta_1-s{b\over 2}, \beta_2-t{b\over
2}}\left[\matrix{\beta_2&-{b\over 2}\cr
\beta_3&\beta_1} \right]\cr
&\quad ={\zG(t b(Q-2\zb_2))\, \zG(1- s b(Q-2\zb_1))\over
\zG(\frac{1\!-\!s}{2}\!+\!t b(\zb_3\!-\!\b_2\!+\!s t \b_1\!-\!s t {b\over 2}) )\, \zG(\frac{t+1}{2}-\!t  b(\zb_2\!+\!\b_3\!-\!st \b_1\!-\!{b\over 2})\!-\!\frac{s\!-\!t}{2} bQ)\, }\,.
}}
The dual fusion matrix elements  ${\tilde F}_{s,t}^L$  are obtained  with $b \to 1/b$.
 All
  these expressions should be considered as
furthermore restricted
by the fusion rules.
The gauge choice is such that e.g., $F^L_{++}=1$
if $\b_1=0$ since the fusion rule leads to $\b_3=\b_2-b/2$, or, if,  $\b_3=Q$ so that  $\b_1-b/2=Q-\b_2$.
The expression for the  matter fundamental  fusion matrix  elements is obtained from \fga\ by analytic continuation
$b^2 \to -b^2\,$ and $ b\b_i \to b e_i$  (so that  $b (\b_1 -t b/2) \to b( e+t b/2)$)
\eqn\fgam{\eqalign{
F_{s,t}^M:= &
F^M_{e_1-s{b\over 2}, e_2-t{b\over
2}}\left[\matrix{e_2&{b\over 2}\cr
e_3&e_1} \right]\cr
&\ ={\zG(t b(2e_2-e_0))\, \zG(1+ s b(e_0-2e_1))\over
\zG(\frac{1\!+\!s}{2}\!-\!t b(e_3\!-\!e_2\!+\!s t e_1\!+\!s t {b\over 2}))\, \zG(\frac{1-t}{2}\!+\!t  b(e_2\!+\!e_3\!-\!s t e_1\!+\!{b\over 2})\!+\!\frac{s\!-\!t}{2} be_0)} \,.
}}
The dual $\tF_{s,t}^M$ is recovered from $F_{-s,-t}^M$ by the change $b\to -1/b$.
For the  choice  of  the chiralities of the three fields  as in \mas\
one has   $F_{s,t}^M=F_{-s,t}^L\,, \ \  \tF_{s,t}^M = \tF_{-s,t}^L$
which implies   \diffr.
Furthermore one needs the
 particular fundamental constants in  \pent{}\  and \pentm{}.  In the Liouville case the constant is  given  by  \FZZ\
\eqn\cmcoefa{\eqalign{
&C^L_{\sigma_3 \pm {b\over 2}\,
\beta_1+{b \over 2}}\left[\matrix{-{b\over 2}&\beta_1 \cr
\sigma_3& \sigma_1} \right]\,
 =-{b^2\,  \sqrt{\L}
\Gamma(1-2b\beta_1)\over  \Gamma(1+(Q-2\beta_1)b)}
\, c^L_{(\mp)}(\sigma_3, \b_1,\sigma_1)
}}
with the last constant written down  in \cmcoef.
The  corresponding matter factor
 (obtained  also as analytic continuation of \cmcoef{}) reads
\eqn\mcoefa{\eqalign{
&C^M_{\bsi_3 \mp {b\over 2}\,,e_1-{b \over 2}}
\left[\matrix{{b\over 2}&e_1 \cr
\bsi_3& \bsi_1} \right]= b^2\, {\sqrt{\LM}\, \Gamma(1-2be_1)\over
 \Gamma(1+(e_0-2e_1)b)}\, c^M_{(\pm)}(\bsi_3, e_1, \bsi_1)
 \cr
}}
with the explicit expression given in \mcoef.
Combining  \cmcoefa, \cmcoef{}\ and  \mcoefa, \mcoef\  the product of the coefficients in the l.h.s. of the $t=t'=-1$ identities in \pent{}, \pentm{}\  reads
\eqn\fuse{\eqalign{
& C^M_{\bsi_3 \,
e_2-{b \over 2}}
\left[\matrix{e_2 &{b\over 2} \cr
\bsi_4& \bsi_2} \right]\ C^L_{\sigma_3 \,
\beta_2+{b \over 2}}\left[\matrix{\beta_2&-{b\over 2} \cr
\sigma_4 & \sigma_2} \right]\cr
& =C^M_{\bsi_3 =\bsi_2+\bar{\delta} {b\over 2}
\,,e_2-{b \over 2}}
\left[\matrix{{b\over 2}&e_2 \cr
\bsi_2& \bsi_4} \right]\ C^L_{\sigma_3=\sigma_2 \, -\delta {b\over 2}\,
\beta_2+{b \over 2}}\left[\matrix{-{b\over 2}&\beta_2 \cr
\sigma_2& \sigma_4} \right]
\cr
&{}\cr
& = - \sqrt{\L\, \LM}\, {\Gamma(\frac{1}{b}(Q-2\b_2-b))\over  \Gamma(\frac{1}{b}(Q-2\b_2+b))}\
c(\b_2) \, c^M_{(-\bar{\delta})}(\bsi_2,e_2, \bsi_4) \,c^L_{(\delta)}(\sigma_2,\b_2, \sigma_4)\,.
}}
The Gamma's in \fuse\ are eliminated by
 the leg factor  normalisation  \leg\
and collecting everything we obtain  the relation \gro{}\ for the normalised
 constants $\hat{C}$
 $$
{1\over \Gamma(b(2\b_3\!-\!Q))}\hat{C}_{\zs_3\,,\zb_3}\!\left[\matrix{\zb_2&\beta_1\cr
\zs_4&\sigma_1} \right]\! =\!\Gamma(\frac{1}{b}(Q\!-\!2\b_2))\Gamma(b(Q\!-\!2\b_1))
{C}^L_{\zs_3\,,\zb_3}\!\left[\matrix{\zb_2&\beta_1\cr
\zs_4&\sigma_1} \right]\!{C}^M_{\bsi_3\,,e_3}\left[\matrix{e_2&e_1\cr
\bsi_4&\bsi_1} \right]
 $$

\appendix{B}{\bf The  equation
in  the nonstandard 
 Liouville gravity}

In \KPa\ another version of the Liouville gravity has been constructed trading the standard matter screening charges
for  non-trivial tachyon interaction terms, with "diagonal" matter screening charges $e_0=1/b-b$.
It had led to  functional equations  with shifts  along the diagonal $e \pm e_0$ and the solutions for the 4-point tachyon bulk correlators  in this model were confirmed by a matrix model based construction
\KPb.
To obtain an equation for the  boundary 3-point correlator with the OPE projected to the diagonal shifts $e \pm e_0$ we cannot follow the derivation of \gro,\grod\ above by linear combinations of matter and Liouville pentagon equations. Instead we
shall  exploit   the  ground ring relations of \KPa,  composing  the individual terms  in these relations. E.g., taking the order
${}^{\s_4}B_{\b_2}^{\s_3}a_-^{\s_2'=\s_3'}a_+^{\s_2}B_{\b_1}^{\s_1}$ in the product of the tachyons with the ground ring generators $a_{\pm}$ 
one obtains (cancelling an overall sign)
\eqn\grodiag{\eqalign{
&   \sqrt{\tM \LM \L }\,
\tilde{c}^M_{(\bar{\delta})}(\bsi_2\!=\!\bar{\s}_3\!-\!\frac{\bar{\delta'} b}{2}\!+\!\frac{\bar{\delta}}{2b}, e_2\!-\!\frac{b}{2}, \bsi_4) \,c(\b_2)
c^M_{(-\bar{ \delta'} )}(\bar{\s}_3\!-\!{ \bar{\delta'}  b\over 2}, e_2, \bsi_4)
c^L_{(\delta')}(\zs_3\!+\!{ \delta'  b\over 2}, \b_2, \sigma_4) \times \cr
&
 \hC_{\s_2
\,,\zb_3}\left[\matrix{\zb_2-{e_0\over 2}&\beta_1\cr
\zs_4&\sigma_1} \right]
 + \sqrt{\tL}\,\tilde{c}^L_{(\delta)}(\sigma_2=\zs_3+\frac{\delta}{ 2b}+ \frac{\delta'  b}{2},
\b_2-\frac{b}{2} , \sigma_4)\,  \hC_{\s_2
\,,\zb_3}\left[\matrix{\zb_2+{e_0\over 2}&\beta_1\cr
\zs_4&\sigma_1} \right]  \cr
&{}\cr
&=
  \sqrt{\LM\tL\tM}\,c^M_{(\bar{\delta'})}(\bsi_3,
    e_1+\frac{1}{2b}, \bsi_1) \,
\tilde{c}(\b_1) \, \tilde{c}^M_{(-\bar{\delta})}(\bar{\s}_2-{\bar{\delta} \over 2b},e_1, \bsi_1) \,\tilde{c}^L_{(-\delta)}(\zs_2-{\delta \over 2b},\b_1, \sigma_1)\ \times \cr
& \hC_{\s_3
\,,\zb_3}\left[\matrix{\zb_2&\beta_1+{e_0\over 2}\cr
\zs_4&\sigma_1} \right]  + \sqrt{\L}\,c^L_{(-\delta')}(\s_3,
\b_1-  \frac{1}{2b}, \sigma_1) \, \hC_{\s_3 \,,\zb_3}\left[\matrix{\zb_2&\beta_1-{e_0\over 2}\cr
\zs_4&\sigma_1} \right] \,.
}}
The opposite order leads to a minus sign for each term
so that the final relation  does not change. We may restrict to diagonal shifts of the boundary labels as well,  taking
 $\delta'=-\delta\,, \ \bar{\delta'}=\bar{\delta}$.  

\appendix{C}{\bf Examples}

\noindent
{\bf Example 1:}   $e_{123}=e_0+b=1/b$
\medskip
The  matter formula \matcob\  reads (set $\LM=1$)
\eqn\matcobex{\eqalign{
&C^M_{\bsi_2 \,,e_0-e_3}\left[\matrix{e_2&e_1\cr
\bsi_3&\bsi_1} \right]=-{{\Pi}_M(e_3,e_2,e_1) \over  2\sin\pi b^2 2\sin\pi 2e_1 b
}\(c^M_{(-)}(\bsi_3,e_3-b/2,\bsi_1)+ c^M_{(+)}(\bsi_3,e_2-b/2,\bsi_2)\)\cr
&={b^2 \G(b^2)\, \prod_i\G(1-2e_ib)\over (2\pi)^2}2(\sin \pi 2e_2 b \, c_1^M+  \sin \pi 2e_3 b \, c_2^M+ \sin \pi 2e_1 b \, c_3^M)\cr
}}
where $c_i^M=2 \cos \pi b(b+2\bsi_i)$. The cyclic symmetry of the 3-point function is explicit.
As a particular example one recovers from \matcob\ the OPE constant  \mcoefa, leading to \mcoef.

We shall
use \matcobex\  for the particular choice of three degenerate matter fields
\eqn\extb{
\b_1=\b_2=\b_3=2b \to e_1 =e_3=b= e_0-e_2\,
}
or any other choice of  two $(+)$ and one $(-)$ chiralities. 
For our example $k=2$ in \ants\  is even and the polynomial  \ants\   in the variables $c_2,c_1$ 
\eqn\extw{\eqalign{
  B(\s_2,\s_1)^{(2;0)} &=
(c_2-c_1)g_-(\s_2,- \frac{b}{2},\s_1)g_-(\s_2, \frac{3 b}{2},\s_1) =(c_2-c_1) P(c_2,c_1)\cr
}}
 is antisymmetric. 
 The polynomial \csympol\  is symmetric in $\s_1,\s_3$ and is proportional to
\eqn\simpol{
P_2:=g_-(\s_2,- \frac{b}{2},\s_1)+g_-(\s_2, \frac{3 b}{2},\s_3)=-\sum_{i=1}^3c_i+c_2\, (1+ 2 \cos \pi 2b^2)\,.
}
Then the Liouville factor   \csymf{}\   reads 
\eqn\nderaba{\eqalign{
&{}^LC_{2b, 2b, 2b}^{\s_3,\s_2 ,\s_1} =\cr
&= {S_b(3b) S_b(2b)\over S_b^2(b)}  {\L^{Q-6b\over 2b}\,\, \Pi_L(2b,2b,2b)    \over     B(\s_1,\s_2)^{(2;0)} \,B(\s_2,\s_3)^{(2;0)}  \, B(\s_3,\s_1)^{(2;0)} }  \det\pmatrix{\tilde{c}_3\,{X}_3&\tilde{c}_2\,{X}_2&\tilde{c}_1\,{X}_1\cr
c_3&c_2&c_1\cr
1&1&1} \, \cr
}}
with $X_3= X_3(c_1,c_2,c_3): = P_3 \, P(c_1,c_2)$. Combining \nderaba\  with  \matcobex\  and
the full prefactor from \prefful\  one obtains the tachyon correlator in this example.
 Note that for the  choice of the chiralities $-\e_1=1=\e_2=\e_3$, 
 the matter correlator \matcobex\   is indeed proportional to  the 
 polynomial $P_3$ in \simpol,  since all $ c_k^M$  are  identified with $c_k$. Similarly
 the choice  of the negative chirality  as $\e_2=-1$ or $\e_3=-1$ leads to the polynomial  $ P_1$ or $P_2$
 respectively. 
\medskip

\noindent
{\bf Example 2:}    $e_{123}=e_0-\frac{1}{b}=-b$
\medskip
The  matter formula \matcob\  reads
\eqn\matcobexd{\eqalign{
&C^M_{\bsi_2 \,,e_0-e_3}\left[\matrix{e_2&e_1\cr
\bsi_3&\bsi_1} \right]={{\Pi}_M(e_3,e_2,e_1) \over  2\sin\pi /b^2 2\sin\pi 2e_1/b}
\(\tilde{c}^M_{(+)}(\bsi_3,e_3+\frac{1}{2b},\bsi_1)+ \tilde{c}^M_{(-)}(\bsi_3,e_2+\frac{1}{2b},\bsi_2)\)\cr
&={1\over b^2} {\G(\frac{1}{b^2}) \prod_i\G(1+\frac{2e_i}{b})\ \over (2\pi)^2}
2(\sin \pi (-\frac{2e_2}{b}) \, \tilde{c}_1^M+  \sin \pi (-\frac{2e_3}{b})\, \tilde{c}_2^M+ \sin \pi (-\frac{2e_1}{b} )\, \tilde{c}_3^M)\,,
}}
where $\tilde{c}_i^M=2 \cos \pi \frac{1}{b}(\frac{1}{b}-2\bsi_i)$. Comparing with \matcobex\ one observes that the symmetry $b \to -1/b$ of the correlator is indeed confirmed.
The matter  correlator \matcobexd\ can be used e.g., to compute the tachyon 3-point function with
 $$\b_3=\b_2=\b_1= b-1/b \Rightarrow  e_1=e_3=-1/b=e_0-e_2\,.$$
The Liouville 3-point function in this case has been given   in \Alex\ and it is
 cast in  a 
  form similar to \nderaba,
\eqn\seempl{
{}^LC_{b-\frac{1}{b}, b-\frac{1}{b}, b-\frac{1}{b}}^{\s_3,\s_2 ,\s_1} ={S_b(\frac{1}{b}) S_b(\frac{3}{b})\over S_b(\frac{2}{b})S_b(\frac{5}{b})}
{\L^{Q-3 e_0\over 2b}\,
\Pi_L(b\!-\!\frac{1}{b},b\!-\!\frac{1}{b},b\!-\!\frac{1}{b})  
 \over  
 ({c}_1-{c}_2)   ({c}_2-{c}_3) ({c}_3-{c}_1)  }
\det\pmatrix{c_3\tilde{X}_3&c_2\tilde{X}_2&c_1\tilde{X}_1\cr
\tilde{c}_3&\tilde{c}_2&\tilde{c}_1\cr
1&1&1}} 
where $\tilde{X}_i$ is the dual ($ b\to 1/b$) of the polynomial  $X_i$ in \nderaba. The duality $b\to 1/b$
transformation of  \seempl,   so that $\b_i=b-1/b \to 1/b-b$,   gives a  new correlator,  which is obtained alternatively from \nderaba\  by reflecting all three boundary fields $ \b_i=2b \to Q-2b=1/b-b$ with the corresponding 2-point reflection amplitudes. 
\medskip

 \listrefs

\bye